\documentclass[twocolumn,twoside,slac_two]{revtex4}
\usepackage{graphicx}
\usepackage{fancyhdr}
\def\pr{{Phys. Rev.}~}
\def\prl{{ Phys. Rev. Lett.}~}
\def\pl{{ Phys. Lett.}~}
\def\np{{ Nucl. Phys.}~}

\begin{document}

\title{Theoretical review on $\sin2\beta(\phi_1)$ from $b \to s$ penguins}


%\author{Y. Lei}
%\affiliation{National Taiwan University, Taipei, Taiwan}
%
\author{Chun-Khiang Chua}
\affiliation{Department of Physics, Chung-Yuan Christian
University, Taiwan 32023, Republic of China}

\begin{abstract}
Recent theoretical results of the standard model expectations on
$\sin2\beta_{\rm eff}$ from penguin-dominated $b \to s$ decays are
briefly reviewed.
\end{abstract}

\maketitle

\thispagestyle{fancy}


\section{Introduction}

Possible New Physics beyond the Standard Model is being
intensively searched via the measurements of time-dependent $CP$
asymmetries in neutral $B$ meson decays into final $CP$
eigenstates defined by
 \begin{eqnarray}
 &&{\Gamma(\overline B(t)\to f)-\Gamma(B(t)\to f)\over
 \Gamma(\overline B(t)\to f)+\Gamma(B(t)\to f)}
 \nonumber\\
 &&={\cal S}_f\sin(\Delta mt)+{\cal A}_f\cos(\Delta mt),
 \end{eqnarray}
where $\Delta m$ is the mass difference of the two neutral $B$
eigenstates, $S_f$ monitors mixing-induced $CP$ asymmetry and
${\cal A}_f$ measures direct $CP$ violation. The $CP$-violating
parameters ${\cal A}_f$ and ${\cal S}_f$ can be expressed as
 \begin{eqnarray}
 {\cal A}_f=-{1-|\lambda_f|^2\over 1+|\lambda_f|^2},
 \qquad
 {\cal S}_f={2\,{\rm Im}\lambda_f\over 1+|\lambda_f|^2},
 \end{eqnarray}
where
 \begin{eqnarray}
 \lambda_f={q_B\over p_B}\,{A(\overline B^0\to f)\over A(B^0\to f)}.
 \end{eqnarray}

\begin{figure}[b]
\centering
\includegraphics[width=80mm]{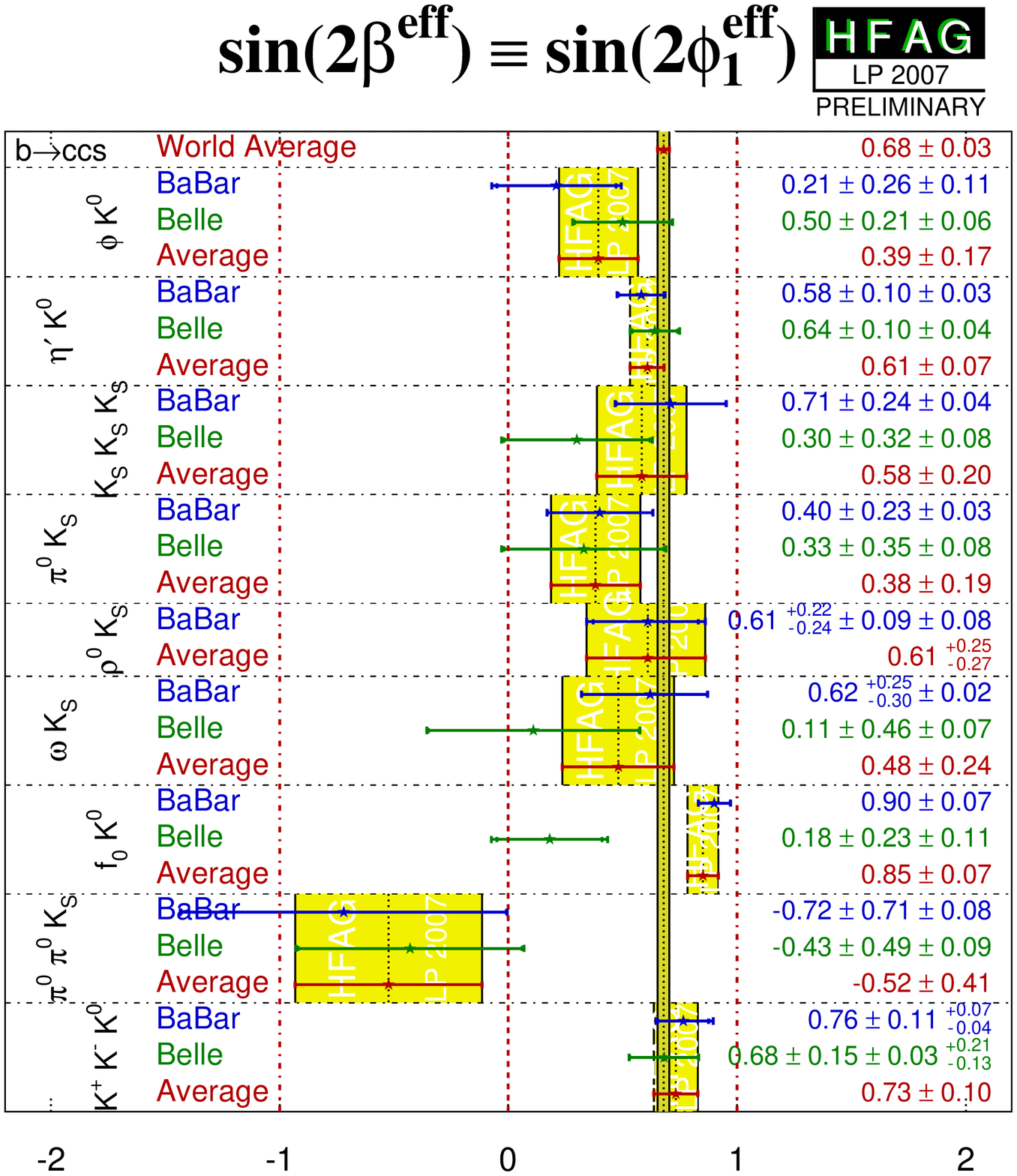}
\caption{Experimental results for $\sin2\beta_{\rm eff}$ from
$b\to s$ penguin decays~\cite{HFAG2008}.} \label{sin2beta}
\end{figure}

In the standard model $\lambda_f\approx \eta_f e^{-2i\beta}$ for
$b\to s$ penguin-dominated or pure penguin modes with $\eta_f=1$
($-1$) for final $CP$-even (odd) states and $\beta($or
$\phi_1)=\arg(-V_{cd}V^*_{cb}/V_{td}V_{tb}^*)$. Therefore, it is
expected in the Standard Model that $-\eta_fS_f\approx \sin
2\beta$ and ${\cal A}_f\approx 0$.

The mixing-induced $CP$ violation in $B$ decays has already been
observed in the golden mode $\overline B{}^0\to J/\psi K_S$ for
several years. The current world average the mixing-induced
asymmetry from tree $b\to c\bar c s$ transition is~\cite{HFAG2008}
 \begin{eqnarray}
 \sin 2\beta=0.681\pm0.025\,.
 \end{eqnarray}
Results of the time-dependent {\it CP}-asymmetries in the $b\to
sq\bar q$ induced two-body decays such as $\overline B{}^0\to
(\phi,\omega,\pi^0,\eta',f_0)K_S$ are shown in Fig.~1 and
2~\cite{HFAG2008}. In the SM, CP asymmetry in all above-mentioned
modes should be equal to ${\cal S}_{J/\psi K}$ with a small
deviation {\it at most} ${\cal O}(0.1)$ \cite{LS}. As discussed in
\cite{LS}, this may originate from the ${\cal O}(\lambda^2)$
truncation and from the subdominant (color-suppressed) tree
contribution to these processes. Since the penguin loop
contributions are sensitive to high virtuality, New Physics beyond
the SM may contribute to ${\cal S}_f$ through the heavy particles
in the loops. In order to detect the signal of New Physics
unambiguously in the penguin $b\to s$ modes, it is of great
importance to examine how much of the deviation of $S_f$ from
${\cal S}_{J/\psi K}$,
 \begin{eqnarray}
 \Delta {\cal S}_f\equiv -\eta_f {\cal S}_f-{\cal S}_{J/\psi K_S},
 \end{eqnarray}
is allowed in the
SM~\cite{LS,Grossman97,Grossman98,Grossman03,Gronau-piK,Gronau-eta'K,GronauRosner,QCDF,CCY2005,pQCD,SCET,Lu08,CCS2005,CCSKKK,SU3KKK}.

\begin{figure}[t]
\centering
\includegraphics[width=80mm]{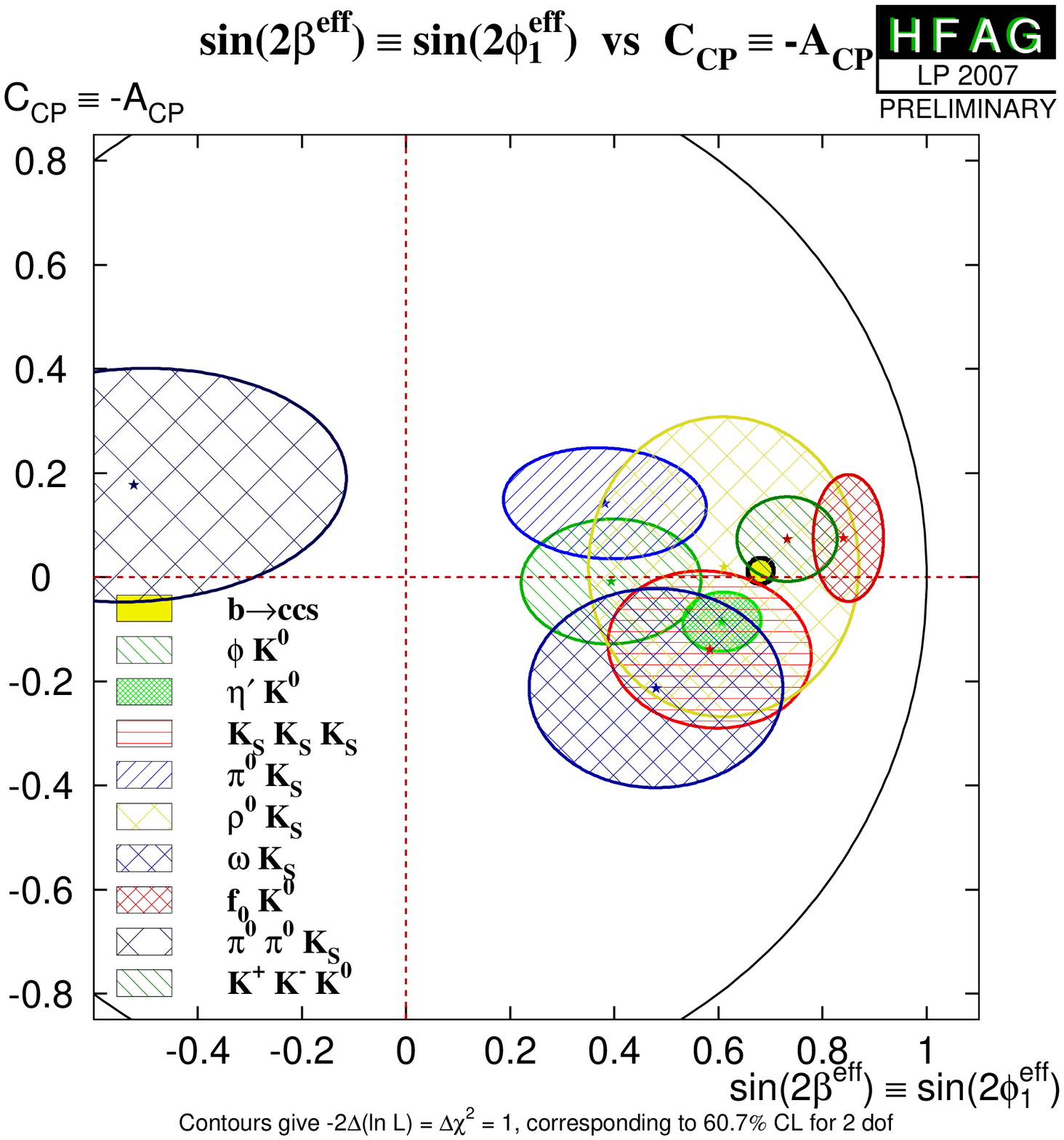}
\caption{Experimental results for $\sin2\beta_{\rm eff}$ and
${\cal A}_f$ from $b\to s$ penguin decays~\cite{HFAG2008}.}
\label{sin2betaC}
\end{figure}

The decay amplitude for the pure penguin or penguin-dominated
charmless $B$ decay in general has the form
 \begin{eqnarray}
 M(\overline B^0\to f) = V_{ub}V_{us}^*F^u+V_{cb}V_{cs}^* F^c
 +V_{tb}V_{ts}^*F^t.
 \end{eqnarray}
Unitarity of the CKM matrix elements leads to
 \begin{eqnarray}
 M(\overline B^0\to f) &=& V_{ub}V_{us}^*A_f^u+V_{cb}V_{cs}^*A_f^c
 \nonumber\\
 &\approx& A\lambda^4R_be^{-i\gamma}A_f^u+A\lambda^2 A_f^c,
 \end{eqnarray}
where we use $A_f^u\equiv F^u-F^t$, $A_f^c\equiv F^c-F^t$ and
$R_b\equiv|V_{ud}V_{ub}/(V_{cd}V_{cb})|=\sqrt{\bar\rho^2+\bar\eta^2}$.
The first term in the above expression is suppressed by a factor
of $\lambda^2$ relative to the second term. For a pure penguin
decay such as $\overline B{}^0\to\phi K_S$, it is naively expected
that $A_f^u$ is in general comparable to $A_f^c$ in magnitude.
Therefore, to a good approximation we have $-\eta_fS_f\approx\sin
2\beta\approx S_{J/\psi K}$. For penguin-dominated modes, such as
$\omega K_S,\rho^0K_S,\pi^0K_S$, $A_f^u$ also receives tree
contributions from the $b\to u\bar u s$ tree operators. Since the
Wilson coefficient for the penguin operator is smaller than the
one for the tree operator, it is possible that $A_f^u$ is larger
than $A_f^c$. As the $b\to u$ term carries a weak phase $\gamma$,
$S_f$ may be subjected to a significant ``tree pollution".

To quantify the deviation, it is known
that~\cite{Gronau,Grossman03}
 \begin{eqnarray} \label{eq:CfSf}
 \Delta {\cal S}_f=2|r_f|\cos 2\beta\sin\gamma\cos\delta_f,
 \,\, {\cal A}_f=2|r_f|\sin\gamma\sin\delta_f,
 \nonumber
 \end{eqnarray}
with $r_f\equiv(V_{ub}V_{us}^* A_f^u)/(V_{cb}V_{cs}^* A_f^c)$ and
$\delta_f\equiv{\rm arg}(A_f^u/A_f^c)$ and only terms up to the
first order in $r_f$ are shown. % in the above equation.
Hence, the magnitude of the $CP$ asymmetry difference $\Delta
{\cal S}_f$ and direct $CP$ violation are both governed by the
size of $A_f^u/A_f^c$.
%
%However,
For the aforementioned penguin-dominated modes, the tree
contribution is color suppressed and, hence, in practice, the
deviation of ${\cal S}_f$ is expected to be small~\cite{LS}. It is
useful to note that $\Delta {\cal S}_f$ is proportional to the
real part of $A^u_f/A^c_f$ as shown in the above equation.

Below I will briefly review the results of the SM expectations on
$\Delta {\cal S}_f$ from the SU(3$)_F$ approach, short-distance
and long-distance calculations.

\section{$\Delta{\cal S}_f$ from the SU(3$)_F$ approach}

I briefly review the underlying reasoning of the SU(3$)_F$
approach (using \cite{Grossman03} as an example) and summarize the
present results. Recent reviews of results obtained from the
SU(3$)_F$ approach can be found in~\cite{SU3,Buchalla:2008jp}.

For a $\Delta S=0$ decay, such as $\overline B{}^0\to f'$ decay,
the decay amplitude is given by
 \begin{equation}
 A(\overline B{}^0\to
 f')=V_{ub}V_{ud}^*B_{f'}^u+V_{cb}V_{cd}^*B_{f'}^c.
 \end{equation}
Note that comparing with the $\Delta S=1$ amplitude, we have $s$
replaced by $d$ in the CKM matrices, resulting an opposite
hierarchy of tree and penguin amplitudes. Hence, ratio of (tree
dominated) $\Delta S=0$ and (penguin dominated) $\Delta S=1$
amplitudes may provide information on $r_f$.

Through SU(3$)_F$ symmetry, one can obtain
 \begin{equation}
 A_f^{u(c)}=\sum_{f'} C_f^{f'} B_{f'}^{u(c)},
 \end{equation}
where $C_f^{f'}$ are some SU(3) Clebsch-Gordan coefficients.
Consequently, a suitable sum of $\overline B{}^0\to f'$ amplitudes
gives
 \begin{eqnarray}
 A'(\overline B{}^0\to f)
 &\equiv&\sum_{f'} C_f^{f'}A(\overline B{}^0\to f')
 \nonumber\\
 &=&V_{ub}V_{ud}^*A_{f}^u+V_{cb}V_{cd}^*A_{f}^c,
 \end{eqnarray}
which is identical to $A(\overline B{}^0\to f)$, except with
$V_{us,cs}$ replaced by $V_{ud,cd}$. Note that $A'(\overline
B{}^0\to f)$ is not a $\Delta S=1$ decay amplitude, but a sum of
several $\Delta S=0$ amplitudes. The absolute value of the ratio
of $A'(\overline B{}^0\to f)$ and $A(\overline B{}^0\to f)$ with a
suitable CKM factor, gives
 \begin{equation}
 \xi_f\equiv\Big|\frac{V_{us} A'(\overline B{}^0\to f)}{V_{ud} A(\overline B{}^0\to
 f)}\Big|
      =\Big|\frac{r_f+V_{us}V_{cd}/V_{ud}V_{cs}}{1+r_f}\Big|,
 \end{equation}
which can be used to constrain $r_f$. There are two comments: (i)
From the above expression, we see that the bound on $r_f$ cannot
be better than $|V_{us}V_{cd}/V_{ud}V_{cs}|={\cal O}(\lambda^2)$.
(ii) If no assumption on phases of $\overline B{}^0\to f'$
amplitudes is made, the above ratio is bounded by
   \begin{equation}
   \xi_f\leq \sum_{f'} |V_{us}/V_{ud}|| C_f^{f'}|
   \sqrt{\frac{{\cal B}(\overline B{}^0\to f')}{{\cal B}(\overline B{}^0\to
   f)}},
   \end{equation}
which is, however, a rather conservative bound. The bounds work
better for modes with less ($\Delta S=0$) $\overline B{}^0\to f'$
modes involved in the sum.

Present results on the bounds are briefly summarized, while more
detail discussions can be found in recent
reviews~\cite{SU3,Buchalla:2008jp}. Bounds on various $\xi_f$ are
found to be: $\xi_{\eta'K_s}<0.116$~\cite{Gronau:2006qh,SU3},
$\xi_{K^+K^-K^0}<1.02$ and $\xi_{K_SK_SK_S}<0.31$~\cite{SU3KKK}.
Other results on $\eta'K_S$ and $\pi^0 K_S$ modes can be found in
\cite{Gronau:2006qh}.
These bounds can be improved by measuring relevant $\Delta S=0$
modes as much as possible. For example, measurements of
$\pi^0\eta^{(\prime)}$ and $\eta^{(\prime)}\eta^{(\prime)}$ rates
can improve the $\xi_{\eta'K_S}$ bound (see \cite{Ford} for recent
update on the data).

\section{$\Delta{\cal S}_f$ from short-distance calculations}

\subsection{$\Delta {\cal S}_f$ in two-body modes}

There are several QCD-based approaches in calculating hadronic $B$
decays~\cite{BBNS,KLS,BFPS}. $\Delta {\cal S}_f$ from calculations
of QCDF~\cite{QCDF,CCY2005}, pQCD~\cite{pQCD},
SCET~\cite{SCET,Lu08} are summarized in Table~1. The QCDF
calculations on $PP$, $VP$ modes are from
\cite{QCDF}~\footnote{Results obtained agree with those in
\cite{CCS2005}.}, while those in $SP$ modes are from
\cite{CCY2005}. The SCET calculations on $PP$ modes are
from~\cite{SCET}, while those on $VP$ modes are from~\cite{Lu08}.
It is interesting to note that (i)~$\Delta {\cal S}_f$ are
predicted to be small and positive in most cases, while
experimental central values for $\Delta {\cal S}_f$ are all
negative, except the one from $f_0 K_S$; (ii)~In most cases, QCDF
and pQCD results agree with each other, since the main difference
of these two approach is the (penguin) annihilation contribution,
which hardly affects $S_f$; (iii)~The SCET results involve some
non-perturbative contributions fitted from data. These
contributions affect $\Delta S_f$. In some modes results different
from other short distance calculations are obtained.

\begin{table}[t]
\begin{center}
\caption{$\Delta {\cal S}_f$ from various short-distance calculations.} %\label{tab:CP}
\begin{tabular}{|l| c c c |c|}
%\hline
\hline$\Delta {\cal S}_f$
       &QCDF %\,\,\,
       &pQCD %\,\,\,
       &SCET %\,\,\,
       &Expt %\,\,\,\,\,\,\,\,
       \\
       \hline
 $\phi K_S$
       & $0.02\pm0.01$
       & $0.03\pm0.03$
       & $0.01$
       & $-0.29\pm0.17$
       \\
 $\omega K_S$
       & $0.13\pm0.08$
       & $0.16^{+0.04}_{-0.07}$
       & $\begin{array}{c}-0.18^{+0.06}_{-0.07}\\0.12\pm{0.03}\end{array}$
       & $-0.20\pm{0.24}$
       \\
 $\rho^0K_S$
       & $-0.08^{+0.08}_{-0.12}$
       & $-0.18^{+0.10}_{-0.07}$
       & $\begin{array}{c}0.17^{+0.05}_{-0.06}\\-0.12^{+0.03}_{-0.04}\end{array}$
       & $-0.07_{-0.27}^{+0.25}$
       \\
 $\eta' K_S$
       & $0.01\pm0.01$
       &
       & $\begin{array}{c}-0.02\pm0.01\\-0.01\pm0.01\end{array}$
       & $-0.07\pm0.08$
       \\
 $\eta K_S$
       & $0.10^{+0.11}_{-0.07}$
       &
       & $\begin{array}{c}-0.03\pm0.17\\+0.07\pm0.14\end{array}$
       &
       \\
 $\pi^0K_S$
       & $0.07^{+0.05}_{-0.04}$
       & $0.06^{+0.02}_{-0.03}$
       & $0.08\pm0.03$
       & $-0.30\pm0.19$
       \\
 $f_0K_S$
       & $0.02\pm0.00$
       &
       &  %\footnotemark[1]
       & $+0.17\pm0.07$
       \\
  $a_0K_S$
       & $0.02\pm0.01$
       &
       &
       &
       \\
  $\bar K^{*0}_0\pi^0$
       & $\begin{array}{c}0.00^{+0.03}_{-0.05}\\0.02^{+0.00}_{-0.02}\end{array}$
       &
       &
       &
       \\
       \hline
       %\hline
\end{tabular}
\end{center}
\end{table}

It is instructive to understand the size and sign of $\Delta {\cal
S}_f$ in the QCDF approach~\cite{QCDF}, for example.  Recall that
$\Delta S_f$ is proportional to the real part of $A^u_f/A^c_f$,
which we shall pay attention to. We follow \cite{QCDF} to denote a
complex number $x$ by $[x]$ if ${\rm Re}(x)>0$. In QCDF the
dominant contributions to $A^u_f/A^c_f$ are basically given
by~\cite{QCDF,BNb}
 \begin{eqnarray}
 \label{eq:QCDF}
 \frac{A^u_{\phi K_S}}{A^c_{\phi K_S}}%\bigg|_{\phi K_S}
 &\!\!\!\!\sim&\frac{[-(a^u_4+r_\chi a^u_6)]}{[-(a^c_4+r_\chi a^c_6)]}\sim\frac{[-P^u]}{[-P^c]},
 \nonumber\\
 \frac{A^u_{\omega K_S}}{A^c_{\omega K_S}}%\bigg|_{\omega K_S}
 &\!\!\!\!\sim&\frac{+[a^u_4-r_\chi a^u_6]+[a^u_2 R]}{+[a^c_4-r_\chi a^c_6]}\sim\frac{+[P^u]+[C]}{+[P^c]},
 \nonumber\\
 \frac{A^u_{\rho K_S}}{A^c_{\rho K_S}}%\bigg|_{\rho K_S}
 &\!\!\!\!\sim&\frac{-[a^u_4-r_\chi a^u_6]+[a^u_2 R]}{-[a^c_4-r_\chi a^c_6]}\sim\frac{-[P^u]+[C]}{-[P^c]},
 \\
 \frac{A^u_{\pi^0 K_S}}{A^c_{\pi^0 K_S}}%\bigg|_{\pi^0 K_S}
 &\!\!\!\!\sim&\frac{[-(a^u_4+r_\chi a^u_6)]+[a^u_2 R']}{[-(a^c_4+r_\chi a^c_6)]}\sim\frac{[-P^u]+[C]}{[-P^c]},
 \nonumber\\
 \frac{A^u_{\eta' K_S}}{A^c_{\eta' K_S}}%\bigg|_{\eta' K_S}
 &\!\!\!\!\sim&\frac{-[-(a^u_4+r_\chi a^u_6)]+[a^u_2 R'']}{-[-(a^c_4+r_\chi a^c_6)]}\sim\frac{[-P^u]-[C]}{[-P^c]},
 \nonumber
 \label{eq:QCDF}
 \end{eqnarray}
where $a^p_i$ are effective Wilson coefficients~\footnote{In
general, we have Re$(a_2)>0$, Re$(a_6)<$Re$(a_4)<0$.}, $r_\chi=
O(1)$ are the chiral factors and $R^{(\prime,\prime\prime)}$ are
(real and positive) ratios of form factors and decay constants.

From Eq.(8), it is clear that $\Delta {\cal S}_f>0$ for $\phi
K_S$, $\omega K_S$, $\pi^0K_S$, since their Re$(A^u_f/A^c_f)$ can
only be positive. Furthermore, due to the cancellation between
$a_4$ and $r_\chi a_6$ in the $\omega K_S$ amplitude, the
corresponding penguin contribution is suppressed. This leads to a
large and positive $\Delta {\cal S}_{\omega K_S}$ as shown in
Table~I. For the cases of $\rho^0 K_S$ and $\eta' K_S$, there are
chances for $\Delta {\cal S}_f$ to be positive or negative. The
different signs in front of $[P]$ in $\rho^0 K_S$ and $\omega K_S$
are originated from the second term of the wave functions $(u\bar
u\pm d\bar d)/\sqrt2$ of $\omega$ and $\rho^0$ in the $\overline
B^0\to \omega$ and $\overline B^0\to\rho^0$ transitions,
respectively. The $[P]$ in $\rho^0 K_S$ is also suppressed as the
one in $\omega K_S$, resulting a negative $\Delta {\cal S}_{\rho^0
K_S}$. On the other hand, $[-P]$ in $\eta' K_S$ is not only
unsuppressed (no cancellation in the $a_4$ and $a_6$ terms), but,
in fact, is further enhanced due to the constructive interference
of various penguin amplitudes~\cite{BNa}. This enhancement is
responsible for the large $\eta' K_S$ rate~\cite{BNa} and also for
the small $\Delta {\cal S}_{\eta' K_S}$~\cite{QCDF,CCS2005}.

\begin{table*}[t]
\caption{Mixing-induced and direct $CP$ asymmetries for various
charmless 3-body $B$ decays~\cite{CCSKKK,Cheng:2008vy}.
Experimental results are taken from
\cite{HFAG2008}.}\label{tab:AS}
%\begin{ruledtabular}
\begin{center}
\begin{tabular}{|l r r r | r r|}
\hline
 Modes
    & ${\cal S}_f\qquad\qquad$
    & $\Delta{\cal S}_f\quad$
    & Expt
    & ${\cal A}_f(\%)\qquad$
    & Expt
    \\
    \colrule
$K^+K^-K_S$
    & $0.728^{+0.001+0.002+0.009}_{-0.002-0.001-0.020}$
    & $0.041^{+0.028}_{-0.033}$ & $0.05\pm0.11$ & $-4.63^{+1.35+0.53+0.40}_{-1.01-0.54-0.34}$
    & $-7\pm8$
    \\
$K_SK_SK_S$
    & $0.719^{+0.000+0.000+0.008}_{-0.000-0.000-0.019}$
    & $0.039^{+0.027}_{-0.032}$ & $-0.10\pm0.20$ & $0.69^{+0.01+0.01+0.05}_{-0.01-0.03-0.07}$
    & $14\pm15$
    \\
$K_S\pi^0\pi^0$
    & $0.729^{+0.000+0.001+0.009}_{-0.000-0.001-0.020}$
    & $0.049^{+0.027}_{-0.032}$
    & $-1.20\pm0.41$
    & $0.28^{+0.09+0.07+0.02}_{-0.06-0.06-0.02}$
    & $-18\pm22$
    \\
$K_S\pi^+\pi^-$
   & $0.718^{+0.001+0.017+0.008}_{-0.001-0.007-0.018}$
   & $0.038^{+0.031}_{-0.032}$
   &
   & $4.94^{+0.03+0.03+0.32}_{-0.02-0.05-0.40}$
   &
   \\
\hline
\end{tabular} %\label{tab:CP}
%\end{ruledtabular}
\end{center}
\end{table*}

\subsection{$\Delta {\cal S}_f$ in $KKK$ modes}

$\overline B {}^0\to K^+K^-K_{S}$ and $\overline B {}^0\to
K_SK_SK_S$ are penguin-dominated and pure penguin decays,
respectively. They are also used to extracted $\sin2\beta_{\rm
eff}$ with results shown in Fig.~1 and 2.

Three-body modes are in general more complicated than two-body
modes. A factorization approach is used to study these $KKK$
modes~\cite{CCSKKK}. For a review on charmless three body modes,
one is referred to \cite{Cheng:2008vy}. Results of CP asymmetries
for these modes are summarized in Table~II.

To study $\Delta {\cal S}_f$ and ${\cal A}_f$, it is crucial to
know the size of the $b\to u$ transition term ($A^u_f$). For the
pure-penguin $K_SK_SK_S$ mode, the smallness of $\Delta {\cal
S}_{K_SK_SK_S}$ and ${\cal A}_{K_SK_SK_S}$ can be easily
understood. For the $K^+K^-K_S$ mode, there is a $b\to u$
transition in the $\langle\overline B {}^0\to K^+
K_S\rangle\otimes \langle 0\to K^-\rangle$ term. It has the
potential of giving large tree pollution to $\Delta {\cal
S}_{K^+K^-K_S}$.
%It requires more efforts to study the size and
%the impact of this term.

It is useful to note that the $K^+K^-K_S$ final state in the $b\to
u$ transition is not $CP$ self-conjugated. This can be easily seen
by noting that the $K^-$ meson from the $\langle\overline B
{}^0\to K^+ K_S\rangle\times \langle 0\to K^-\rangle$ term is
produced from the virtual $W^-$ meson. Therefore, the $CP$
conjugated term, $\langle\overline B {}^0\to K^- K_S\rangle\times
\langle 0\to K^+\rangle$ is missing in the weak decay amplitude.
Hence, the $b\to u$ transition term should contribute to both
$CP$-even and $CP$-odd configurations with similar strength.
Consequently, information in the $CP$-odd part can be used to
constrain its size and impact on $\Delta {\cal S}_f$ and ${\cal
A}_f$. Indeed, it is found that the $CP$-odd part is highly
dominated by $\phi K_S$, where other contributions (at
$m_{K^+K^-}\not= m_{\phi}$) are highly suppressed~\cite{HFAG2008}.
Since the $\langle\overline B {}^0\to K^+ K_S\rangle\times \langle
0\to K^-\rangle$ term favors a large $m_{K^+K^-}$ region, which is
clearly separated from the $\phi$-resonance region, the result of
the $CP$-odd configuration strongly constrains the contribution
from this $b\to u$ transition term. Therefore, the tree pollution
is constrained and the $\Delta {\cal S}_{K^+K^-K_{S}}$ should not
be large.

\section{FSI contributions to $\Delta {\cal S}_f$}

\begin{table*}[t]
\begin{center}
\caption{Direct CP asymmetry parameter ${\cal A}_f$ and the
mixing-induced CP parameter $\Delta {\cal S}_f^{SD+LD}$ for
various modes. The first and second theoretical errors correspond
to the SD and LD ones, respectively~\cite{CCS2005}.}
\label{tab:Sf}
%\begin{ruledtabular}
\begin{tabular}{|l| r c c |r c c|}
\hline%\hline
      &  \multicolumn{3}{c|}{$\Delta {\cal S}_f$}
      &   \multicolumn{3}{c|}{${\cal A}_f(\%)$}
      \\
      \cline{2-4}
      \cline{5-7}
\raisebox{2.0ex}[0cm][0cm]{Final State} & SD & SD+LD & Expt & SD &
SD+LD &Expt
      \\
      \hline
 $\phi K_S$
       & $0.02^{+0.01}_{-0.02}$
       & $0.04^{+0.01+0.01}_{-0.02-0.02}$
       & $-0.29\pm0.17$
      & $0.8^{+0.5}_{-0.2}$
      & $-2.3^{+0.9+2.2}_{-1.0-5.1}$
      & $1\pm12$
      \\
 $\omega K_S$
       & $0.12^{+0.06}_{-0.05}$
       & $0.02^{+0.03+0.03}_{-0.04-0.02}$
       & $-0.20\pm{0.24}$
      & $-6.8^{+2.4}_{-4.0}$
      & $-13.5^{+3.5+2.4}_{-5.7-1.5}$
      & $20\pm19$
      \\
 $\rho^0K_S$
       & $-0.08^{+0.03}_{-0.10}$
       & $-0.04^{+0.07+0.10}_{-0.10-0.12}$
       & $-0.07^{+0.25}_{-0.27}$
      & $7.8^{+4.5}_{-2.0}$
      & $48.9^{+15.8+5.8}_{-13.7-12.5}$
      & $-2\pm29$
      \\
 $\eta' K_S$
       & $0.01^{+0.01}_{-0.02}$
       & $0.00^{+0.01+0.00}_{-0.02-0.00}$
       & $-0.07\pm0.08$
      & $1.7^{+0.4}_{-0.3}$
      & $2.1^{+0.2+0.1}_{-0.5-0.4}$
      & $9\pm6$
      \\
 $\eta K_S$
       & $0.07^{+0.03}_{-0.03}$
       & $0.07^{+0.03+0.00}_{-0.03-0.01}$
       & $-$
      & $-5.7^{+2.0}_{-5.5}$
      & $-3.9^{+1.8+2.5}_{-5.0-1.6}$
      & $-$
      \\
 $\pi^0K_S$
       & $0.06^{+0.03}_{-0.03}$
       & $0.04^{+0.01+0.02}_{-0.02-0.02}$
       & $-0.30\pm0.19$
      & $-3.2^{+1.1}_{-2.3}$
      & $3.7^{+1.9+1.7}_{-1.6-1.7}$
      & $-14\pm11$
      \\
 \hline%\hline
\end{tabular}
\end{center}
%\end{ruledtabular}
\end{table*}

\begin{figure}[b]
\centering
\includegraphics[width=60mm]{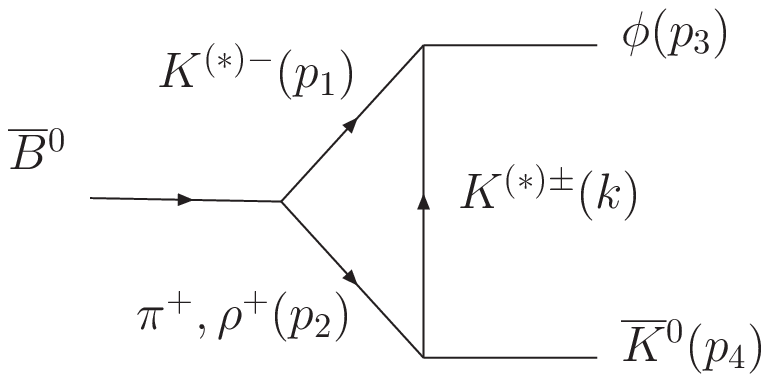}
\includegraphics[width=60mm]{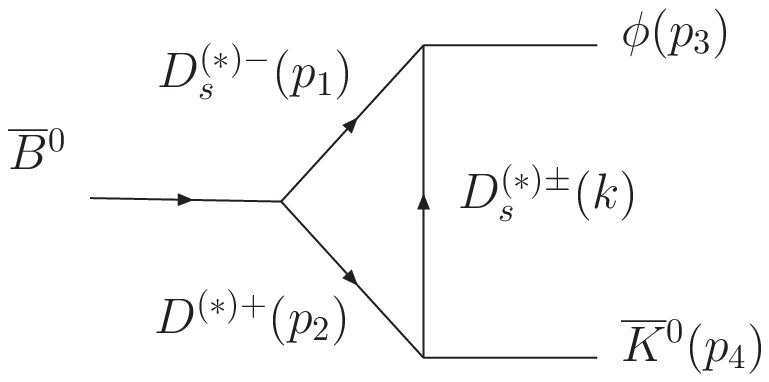}
\caption{Final-state rescattering contributions to the $\overline
B{}^0 \to\phi \overline K{}^0$ decay.} \label{example_figure}
\end{figure}

It was realized recently that long distance FSI may play
indispensable role in $B$ decays~\cite{CCS}. The possibility of
final-state interactions in bringing in possible tree pollution
sources to ${\cal S}_f$ are considered in~\cite{CCS2005}. Both
$A_f^u$ and $A_f^c$ will receive long-distance tree and penguin
contributions from rescattering of some intermediate states. In
particular, there may be some dynamical enhancement on light
$u$-quark loop. If tree contributions to $A^u_f$ are sizable, then
final-state rescattering will have the potential of pushing $S_f$
away from the naive expectation. Take the penguin-dominated decay
$\overline B^0\to \omega \overline K^0$ as an illustration. It can
proceed through the weak decay $\overline B^0\to K^{*-}\pi^+$
followed by the rescattering $K^{*-}\pi^+\to \omega \overline
K^0$. The tree contribution to $\overline B^0\to K^{*-}\pi^+$,
which is color allowed, turns out to be comparable to the penguin
one because of the absence of the chiral enhancement characterized
by the $a_6$ penguin term. Consequently, even within the framework
of the SM, final-state rescattering may provide a mechanism of
tree pollution to ${\cal S}_f$. By the same token, we note that
although $\overline B^0\to \phi \overline K^0$ is a pure penguin
process at short distances, it does receive tree contributions via
long-distance rescattering. Note that in addition to these
charmless final states contributions, there are also contributions
from charmful $D_s^{(*)} D^{(*)}$ final states, see Fig.~3. These
final-state rescatterings provide the long-distance $u$- and
$c$-penguin contributions.

An updated version~\cite{Chua:2006hr} of results in \cite{CCS2005}
are shown in Table~III. Several comments are in order. (i) $\phi
K_S$ and $\eta' K_S$ are the theoretical and experimental cleanest
modes for measuring $\sin2\beta_{\rm eff}$ in these penguin modes.
The constructive interference behavior of penguins in the $\eta'
K_S$ mode is still hold in the LD case, resulting a tiny $\Delta
{\cal S}_{\eta'K_S}$. (ii) Tree pollutions in $\omega K_S$ and
$\rho^0 K_S$ are diluted due to the LD $c$-penguin contributions.
(iii) In general, in this approach, the main contributions to
decay amplitudes are charming-penguin like and do not sizably
affect ${\cal S}_f$.

Recent measurements on $K\pi$ direct CP violations show a more
than 5~$\sigma$ deviation (known as the $K\pi$ puzzle) between
${\cal A}(B^-\to K^-\pi^0)$ and ${\cal A}(\overline B{}^0\to
K^-\pi^+)$~\cite{HFAG2008} . The data indicates the needs of other
sub-leading contributions, such as long distance FSI and charming
penguins and so on (see, for
example~\cite{Chua:2007cm,CharmingP}). It is found that in cases
where the $K\pi$ direct CP data are reproduced, these sub-leading
contributions do not sizably affect the magnitudes of $\Delta
S_f$~\cite{Chua:2007cm}, but some of the signs are different from
the short-distance expectations~\cite{CharmingP}.

\section{Conclusions}

Various theoretical approaches and results on $\Delta {\cal S}_f$
are briefly reviewed. Considerable progress has been made.
From these results we see that the prediction on signs of $\Delta
{\cal S}_f$ are more or less fluctuating and may be subjected to
change when more hadronic contributions are taken into account, on
the contrary, the predictions on the sizes of $\Delta {\cal S}_f$
should be more robust. Since the predictions on sizes of $\Delta
{\cal S}_f$, which are not sizable in most cases, have better
agreement among various approaches. At the same time for modes
with small $\Delta {\cal S}_f$ ($\leq 5\%$), we do not expect
sizable direct $CP$ violations. Measurements on direct $CP$
violations, some $\Delta S=0$ rates and three-body rates and
spectra can provide useful information that can be used to improve
our theoretical predictions on $\Delta {\cal S}_f$.
To further improve the theoretical accuracy more works are needed
to effectively reduce the hadronic uncertainties.

\begin{acknowledgments}
I am grateful to the organizers of FPCP2008 for inviting me to the
exciting conference and to Hai-Yang Cheng and  Amarjit Soni for
very fruitful collaboration.
\end{acknowledgments}

\bigskip % extra skip inserted

\end{document}